\documentclass{ws-p8-50x6-00}
\usepackage{xspace}
\newcounter{enumct}
\newenvironment{Enumerate}{\begin{list}{\arabic{enumct}.}%
{\usecounter{enumct}\setlength{\topsep}{0.2mm}%
\setlength{\partopsep}{0.2mm}\setlength{\itemsep}{0.2mm}%
\setlength{\parsep}{0.2mm}}}{\end{list}}
\newcommand{\epem   }{\ensuremath{\mathrm{e}^+\mathrm{e}^-}\xspace}
\newcommand{\aem    }{\ensuremath{\alpha_\mathrm{em}}\xspace}
\newcommand{\ccbar  }{\ensuremath{\mathrm{c\bar{c}}}\xspace}
\newcommand{\qqbar  }{\ensuremath{\mathrm{q\bar{q}}}\xspace}
\newcommand{\invpb  }{\ensuremath{\mathrm{pb}^{-1}}\xspace}
\newcommand{\qsq    }{\ensuremath{Q^{2}}\xspace}

\newcommand{\qzm    }{\ensuremath{\langle \qsq \rangle}\xspace}

\newcommand{\ft     }{\ensuremath{F_{2}^{\gamma}}\xspace}
\newcommand{\ftc    }{\ensuremath{F_{2,\mathrm{c}}^{\gamma}}\xspace}
\newcommand{\ftcxq  }{\ensuremath{\ftc(x,\qsq)}\xspace}
\newcommand{\ftcxqa }{\ensuremath{\ftc(x,\qzm)}\xspace}
\newcommand{\ftcxqan}{\ensuremath{\ftcxqa/\aem}\xspace}

\newcommand{\pT     }{\ensuremath{p_{\mathrm{T}}}\xspace}

\newcommand{\dz     }{\ensuremath{{\mathrm{D}^{0}}}\xspace}
\newcommand{\ds     }{\ensuremath{{\mathrm{D}^{\star}}}\xspace}

\newcommand{\ttag   }{\ensuremath{\theta_{\rm tag}}\xspace}
\newcommand{\gev    }{\ensuremath{\mathrm{GeV}}\xspace}
\newcommand{\gevsq  }{\ensuremath{\mathrm{GeV}^2}\xspace}

\newcommand{\eb     }{\ensuremath{E_\mathrm{b}}\xspace}
\newcommand{\etag   }{\ensuremath{E_\mathrm{tag}}\xspace}
\newcommand{\Wvis   }{\ensuremath{W_{\mathrm{vis}}}\xspace}

\newcommand{\etads  }{\ensuremath{\eta^{\ds}}\xspace}
\newcommand{\etadsa }{\ensuremath{\vert\etads\vert}\xspace}
\newcommand{\ptds   }{\ensuremath{\pT^{\ds}}\xspace}

\newcommand{\xtds   }{\ensuremath{x_{\mathrm{T}}^{\ds}}\xspace}
\newcommand{\sigdsr }{\ensuremath{\sigma^{\ds}}\xspace}

\newcommand{\sigcc  }{\ensuremath{\sigma(\epem\to\epem\,\ccbar\,X)}\xspace}

\newcommand{\Nall   }{\ensuremath{60.3 \pm 10.3}\xspace}

\newcommand{\Slow   }{\ensuremath{ 4.7 \pm  1.3 \pm 0.9 }\xspace}
\newcommand{\Shig   }{\ensuremath{ 3.0 \pm  0.9 \pm 0.4}\xspace}

\begin{document}

\title{Measurement of the Charm 
 Structure Function of the Photon at LEP\footnotemark}

\footnotetext{\hspace{3.7mm}$^\ast$Presented at the International Conference on
       the Structure and Interactions of the Photon, including 
       the 14th International Workshop on Photon-Photon Collisions,
       Ascona, Switzerland, 2-7 September, 2001.}

\author{\'A. Csilling}

\address{CERN, 
CH-1211, Geneva, Switzerland\footnote{ On leave of absence from 
KFKI Research Institute for Particle and Nuclear Physics,
H-1525 Budapest, P.O.Box 49, Hungary}\\ 
E-mail: Akos.Csilling@cern.ch}


\maketitle

\abstracts{
 Charm production is studied in deep-inelastic electron-photon
 scattering using OPAL data at  \epem centre-of-mass energies
 from 183 to 209~GeV. 
 Charm quarks are identified by exclusive reconstruction
 of \ds mesons. 
 The cross-section of \ds production is measured in a restricted
 kinematic region, and then extrapolated to
 the total charm production cross-section and
 the charm structure function of the photon.
 For $x>0.1$ the measurement is well described by 
 Monte Carlo models and perturbative QCD calculations
 but for $x<0.1$ the predictions
 are lower than the data both in the directly measured region 
 and after the extrapolation.
}

\section{Introduction}
\label{sec:intro}
 The charm component of the photon structure function, \ftc, has been measured 
 by OPAL at LEP2 by applying the well established method  of exclusive
 \ds reconstruction to  deep-inelastic electron-photon scattering
 events.
 The determination of \ftc exploits the fact that the differential 
 cross-section as function of $\qsq$ and Bjorken $x$ 
 is proportional to \ftcxq.~\cite{NIS-9904}
 \par
 Due to the large scale established by their masses, the contribution to
 \ft from charm quarks can be calculated in perturbative QCD, and 
 predictions have been evaluated~\cite{LAE-9401LAE-9602} at 
 next-to-leading order (NLO) accuracy.
 \ftc receives 
 contributions from the point-like and hadron-like components of the 
 photon structure,
 with the hadron-like component dominating at very low values of $x$ and 
 the point-like part accounting for most of \ftc for $x>0.1$.
 \par
 The preliminary results presented here~\cite{OPALPN490} 
 extend the earlier measurement~\cite{OPALPR294} of \ftc 
 using basically the same analysis strategy.
 It is based on 654.1~\invpb of data for \epem centre-of-mass energies 
 from 183 to 209~GeV, recorded by the OPAL experiment
  in the years 1997--2000.
%
%
\section{Data selection}
\label{sec:MC}
 The most important 
 cuts used in the selection of deep-inelastic electron-photon
 scattering events containing a  \ds are  summarised below.
%
\begin{Enumerate} 
 \item 
       An electron candidate must be present with 
       an energy $\etag>0.5\eb$ and a 
       polar angle in
       the ranges $33 <\ttag < 55$~mrad (SW) or $60 <\ttag < 120$~mrad (FD),
       corresponding to  $5<\qsq<100$~\gevsq.
 \item Double-tag events are eliminated by requiring that 
       the sum of all energies in the SW and FD detectors
       opposite to the tag are below 0.25\eb
 \item An exclusively reconstructed \ds candidate must be present with
       a transverse momentum $\ptds>1(3)$~\gev for SW(FD)-tagged events and
       a pseudorapidity $\etadsa<1.5$.
       The \ds meson must decay into $\dz\pi$ 
       with the \dz  decaying into the charged particles
       ${\rm K}\pi$ or ${\rm K}\pi\pi\pi$. 
\end{Enumerate}
%
%
\begin{figure}[tbp]
\begin{center}
\epsfxsize=0.52\linewidth
\epsfbox{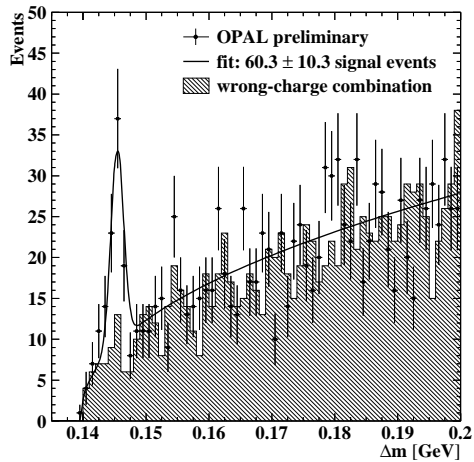}
\caption{Distribution of the difference between the  
         \ds and \dz candidate masses.
         The data are shown as points with statistical errors, while 
         the histogram represents the combinatorial background
         estimated using events with wrong-charge 
         combinations for the decay products of the \ds mesons.
         The curve is the result of the fit to the data.
        }\label{fig:fig02}
\end{center}
\end{figure}
%
 Figure~\ref{fig:fig02} shows the difference between the \ds and 
 \dz candidate masses for both decay channels combined.
 A clear peak is observed around 0.145 GeV, the 
 mass difference between the \ds and the \dz mesons.
 An unbinned maximum likelihood fit to this distribution
 gives \Nall signal
 events above the combinatorial background from deep-inelastic 
 electron-photon scattering events $\epem\to\epem\qqbar$ with
 q=uds.
 The expected background  from all other 
 processes that potentially contain \ds mesons in the final state is
 found to be negligible using Monte Carlo simulations.
 \par
%
%
\begin{figure}[tbp]
\begin{center}
\epsfxsize=0.66\linewidth
\epsfbox{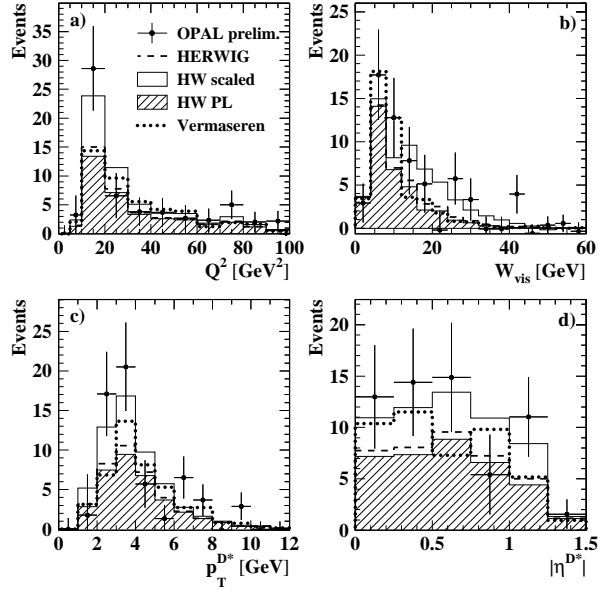}
\caption{Data distributions compared to the HERWIG and Vermaseren predictions.
         For HERWIG several predictions are shown:
         the full prediction, the point-like component alone (HW PL),
         and a superposition of the HERWIG point-like prediction together 
         with a scaled hadron-like prediction, denoted by HW scaled.
        }\label{fig:fig03}
\end{center}
\end{figure}
%
 \par
 Figure~\ref{fig:fig03} shows the distributions of two global event
 quantities, \qsq and \Wvis, and two variables related to the kinematics
 of the \ds candidates, \ptds and \etadsa.
 The data are compared to the absolute predictions of the 
 HERWIG6.1~\cite{MAR-9201} and Vermaseren~\cite{VER-7901VER-8301} 
 leading order Monte Carlo programs.
 \par
\begin{figure}[tbp]
\begin{center}
\epsfxsize=0.65\linewidth
\epsfbox{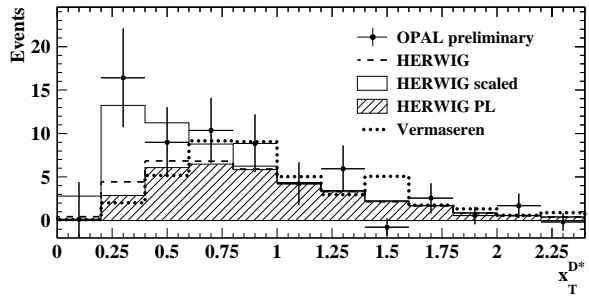}
\caption{The measured \xtds distribution compared to the 
         predictions described in the caption of 
	 Figure~\protect\ref{fig:fig03}.
        }\label{fig:fig04}
\end{center}
\end{figure}

 To get a better description of the data, the 
 hadron-like component of the HERWIG prediction has been 
 fitted to the $\xtds=2\ptds/\Wvis$ distribution shown in 
 Figure~\ref{fig:fig04}  while keeping the point-like part fixed,
 resulting in a scale factor of $6.6 \pm 2.7$.
%
%
%
 There are several possible sources for this difference.
 The NLO prediction itself has a significant uncertainty 
 due to variations of the charm quark mass and the renormalisation 
 and factorisation scales, 
 the gluon distribution of the photon has large experimental errors, and 
 uncertainties in the shape and the modelling of the \ptds distribution
 can change the efficiency for the selected events.
 \par
 This scaled HERWIG prediction, also shown
 in Figure~\ref{fig:fig03}, is used to estimate the 
 signal selection efficiency. The difference between the 
 results obtained with the scaled and the original HERWIG 
 models is taken into account as a systematic uncertainty.
%
%
\section{Results}
%
%
\begin{table}[bp]
\caption{The cross-section \sigdsr measured in the restricted region,
         compared to Monte Carlo predictions.
         The numbers in parentheses refer to the point-like and
         hadron like components.
        }\label{tab:result}
\begin{center}
\small
\begin{tabular}{|c|c|c|}
\hline
\multicolumn{1}{|c|}{}&$0.0014<x<0.1$&$0.1<x<0.87$\\\hline
OPAL          &              \Slow &              \Shig  \\
HERWIG        & 1.02 (0.65 + 0.37) &  2.05 (2.02 + 0.03) \\
HW scaled     & 3.10 (0.65 + 2.45) &  2.23 (2.02 + 0.21) \\
Vermaseren    &               0.84 &                2.81 \\
\hline
\end{tabular}
\end{center}\end{table}
 Table~\ref{tab:result} summarises the cross-section of \ds production 
 in deep-inelastic electron-photon scattering 
 measured in the kinematic region defined by the 
 event selection.
%
\begin{figure}[tbp]
\begin{center}
\epsfxsize=0.65\linewidth
\epsfbox{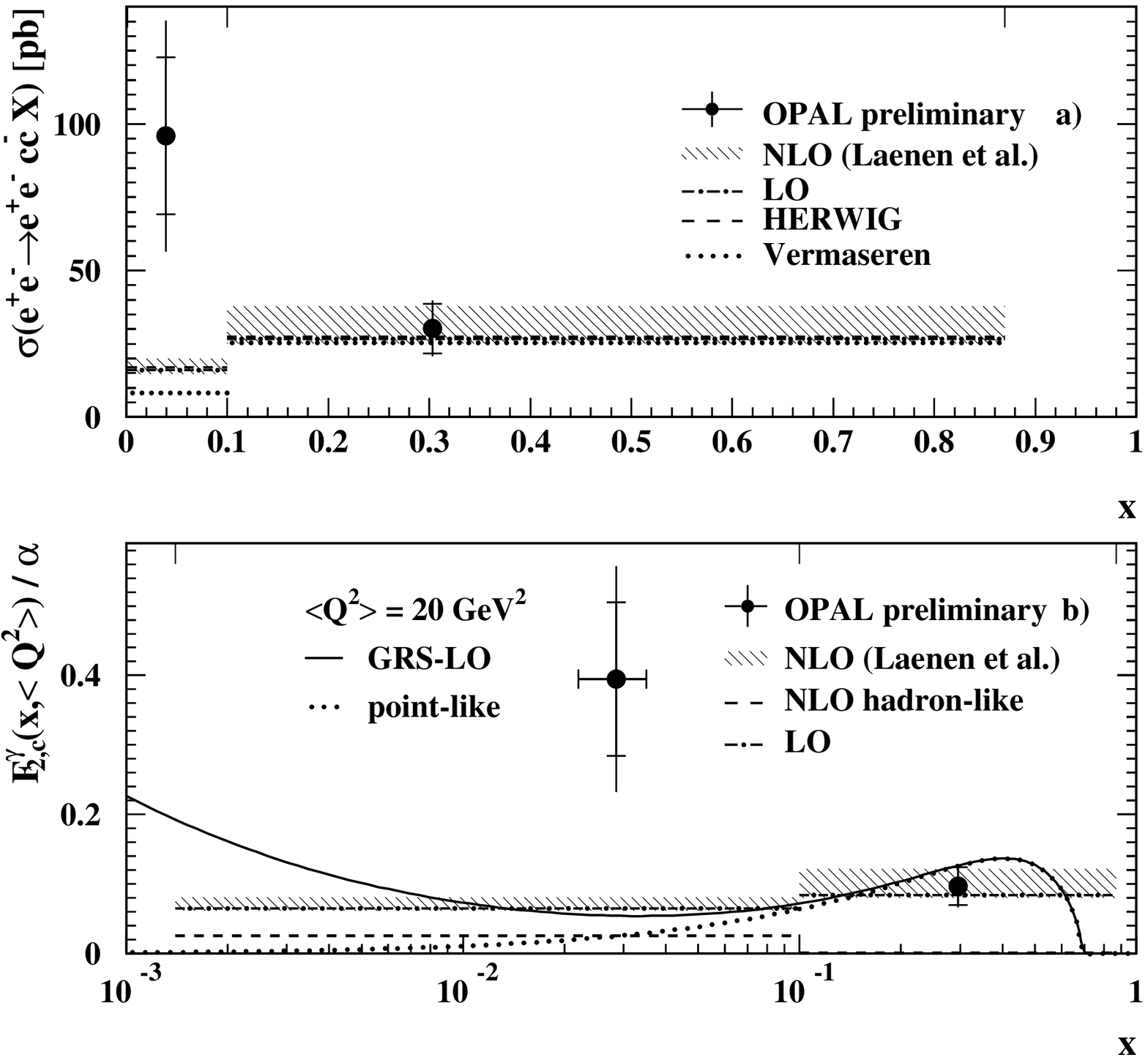}
\caption{Preliminary OPAL results for a) \sigcc, with $5<\qsq<100$~\gevsq and
         b) \ftcxqan.
         The band for the NLO calculation~\protect\cite{LAE-9401LAE-9602}
         indicates the theoretical uncertainties. 
        }\label{fig:fig05}
\end{center}
\end{figure}
%
%
 The total 
 cross-section for \ccbar production in deep-inelastic electron-photon
 scattering,  
 shown in Figure~\ref{fig:fig05}a), 
 is the result of an extrapolation to the whole kinematic region
 using the HERWIG scaled model.
 The value of the charm structure function of the photon, \ftcxqan,
 averaged over the corresponding bin in $x$, 
 shown in Figure~\ref{fig:fig05}b), is obtained using the ratio 
 $\ftcxqan/\sigcc$ given by 
 the NLO calculation.~\cite{LAE-9401LAE-9602}
\par
 All models and predictions shown in Figure~\ref{fig:fig05}
 are in good agreement with the measurement
 for $x>0.1$, where the purely perturbative 
 point-like process is dominant and
 both the experimental and the theoretical uncertainties are
 moderate. 
\par
 On the other hand, for $x<0.1$ the measurement lies more than two
 standard deviations above the predictions, 
 which contain an approximately equal
 contribution from the point-like and hadron-like processes.
 This difference is also observed in the comparisons to the Monte Carlo models
 in the directly measured kinematic region,
 where the extrapolation uncertainties are avoided. Unfortunately, no
 theoretical calculations are available for comparison in the restricted
 phase-space.
\par
%
%
%
%
%

\end{document}